# Towards a multi-physics multi-scale approach of deep geothermal exploration

M. Darnet[1], S. Vedrine[1], F. Bretaudeau[1], S. Marc[1], L. Capar[1], J. PWAVODI[2], G. Marquis[2], A. Montegud[3], J. Formento[3], V. Maurer[4], C. Glass[4], E. Delmais[4], A. Genter[4]

1 BRGM; 2 ITES; 3 CGG; 4 ES-Géothermie

## Summary

A wide range of geophysical methods is used for the exploration of deep geothermal resources. It aims at characterizing the deep fractured network and its capacity for fluid/heat extraction. This relies however on the capacity of geophysical techniques to 1) image the geometry of the fractured network but also 2) characterize the petro-physical properties of the fracture network and matrix. The challenge is however that the geophysical inverse problem is ill-posed and multi-scale.

To overcome these challenges, we propose here to take a multi-physics and multi-scale approach of the geophysical/petro-physical inverse problem. To do so, we are developing iteratively petro-physical and geophysical models that can explain the observables at the different scales. In this paper, we report out the results of the first iteration phase of this research project that consists in building an initial petro-physical model from well logs and prior geological knowledge, and single-domain inversion of geophysical data. We applied this methodology to the Upper Rhine Graben where a wealth of knowledge and dataset on deep fractured formations are available.

**Towards a multi-physics multi-scale approach of deep geothermal exploration**

**Introduction**

Electricity and heat production using Enhanced Geothermal System (EGS) technology rely on our knowledge and prediction capacity of hydrothermal fluid property, mainly temperature and flow rate. Most of hydrothermal fluid properties are obtained from post-drilling phases, whereas proxy information is hard to access in an economically viable way. The main challenge is then to develop methods to access these properties at the early stage of the exploration phase.

A wide range of geophysical methods (active/passive seismics, active/passive electromagnetics, gravity, magnetics) is used at the exploration stage to characterize deep fractured network and its capacity for fluid/heat extraction. This relies however on the capacity of geophysical techniques to 1) image the geometry of the fractured network but also 2) characterize the petro-physical properties of the fracture network and matrix. Allo et al. (2021) showed in the particular example of Dogger carbonates that the petro-physical inversion of seismic data can provide reliable information on the transfer properties of geothermal fluids. It also showed that the inverse problem is ill-posed (several models can explain the same data set) and multi-scale (from core sample, to logs/well, to seismic scale), as demonstrated by Thomas et al. (2023).

To overcome these challenges, we propose to take a multi-physics and multi-scale approach of the geophysical/petro-physical inverse problem. To do so, we are developing iteratively petro-physical and geophysical models that can explain the observables at the different scales. We selected the Upper Rhine Graben (URG) as a demonstrator as for about thirty years, it has been a main target for research on geothermal exploration in deep fractured formations and provides a wealth of geophysical and geological data. In this paper, we report out the results of the initial phase of the integration, namely:
1. Petro-physical modelling of the well logs/core data
2. AI-based inversion of 2D seismic data
3. Inversion of 3D Controlled-Source Electro-Magnetic (CSEM) data

**Petro-physical modelling of the well logs/core data**

From the 64 wells located in the study area, only three have a suite of logs (gamma ray, neutron porosity, density, resistivity and sonic) suitable for quantitative reservoir characterization. A detailed joint petrophysical and rock physics analysis has been performed on all wells to estimate mineral volumes and porosity logs displayed in Figure 1. To help control the quality of those volumes, a rock physics-based approach is used to make the link between the elastic logs predicted and the mineral volumes and porosity logs. The match achieved between predicted and measured elastic logs underlines the importance of running a joint analysis to produce a consistent set of rock volume, validated by the core measurements available (Aichholzer et al. 2019).

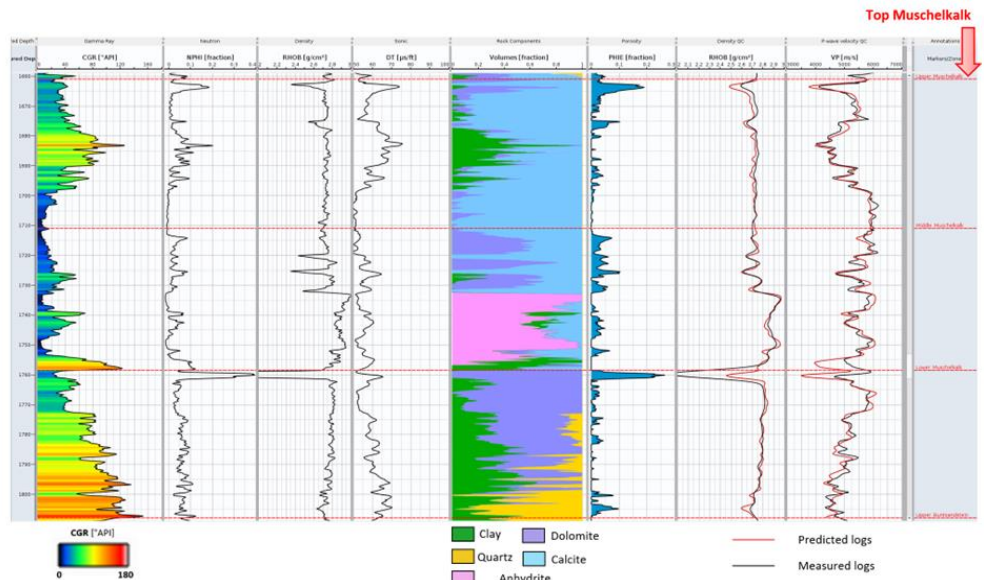

*Figure 1 Petrophysical analysis at one of the wells (GRT1). Mineral volumes (track 5) and porosity (track 6) are computed from petrophysical logs (tracks 1 to 4) using multi-linear regressions. Quality controls include core measurements and elastic log predictions from rock physics models (red curves in tracks 7 and 8).*

Two rock-physics models (RPM) were created and calibrated on the target intervals: Muschelkalk and Buntsandstein. A clay-rich carbonate RPM was used for the Muschelkalk formation considering 5 main minerals volume (calcite, dolomite, anhydrite, clay and quartz) when the Buntsandstein formation can be modelled using a cemented sandstone RPM with 2 main minerals (quartz and clay). These models are crucial to get realistic elastic responses during our synthetic well catalog workflow.

**AI-based inversion of 2D seismic data**

Despite the lack of representative well data in the area, recently introduced theory-guided techniques based on rock physics models can help generate catalogues of pseudo-logs representative of geologic variations (Allo et al. 2021). The same approach was applied on the Upper Rhine Graben dataset by generating hundreds of pseudo-wells considering main variations of the rock properties based on the observed geological variability of the considered formation. Six key 2D lines were selected in the area Rittershoffen (Figure 2) and reprocessed in order to get amplitude preserved suitable for quantitative interpretation. Over the 6 lines, the 2 more recent ones got reasonable offset range to derive angle stacks for elastic property estimations.

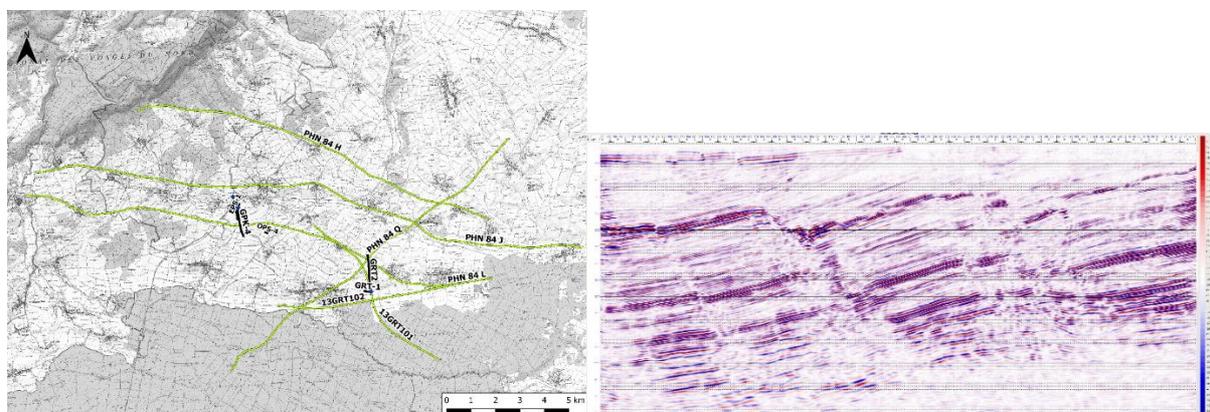

*Figure 2 Map locations with the 2D lines selected for the study and the associated reprocessed 2D line 13GRT102.*

Synthetic seismic traces are generated within the same frequency spectrum of the 2D lines and then used as a training dataset for a deep neural network. This methodology was successfully applied on the geothermal Dogger formation in Paris basin and in this study, the objective is to assess the benefit of using an additional information such as 3D CSEM data in the training set. Several scenarios will be considered in this R&D project trying to quantify the impact of the added information in this ML inversion approach. One of the main challenges is to make the sure of the coherency of the two measurements (seismic and CSEM) regarding the rock property response (porosity or volumes) as well as the scale effects with a difference of resolution between the datasets. The aforementioned study of the petro-physical relationships at core and borehole scales will provide important intial constraints. In addition, a synthetic seismic and CSEM resolution study will be performed to check the validity of these relationships at larger scales.

**Inversion of 3D CSEM data**

To complement seismic data with resistivity data, we used 3D CSEM data from a field survey conducted in the area in 2020. Figure 3 shows the locations of the electrical dipoles and receiver locations. Eight transmission frequencies between 0.03125 and 252 Hz were selected based on electric and magnetic noise information recorded during previous MT campaigns in the area, as well as estimation of investigation depth and sensitivity studies derived from prior resistivity information from well data. An initial resistivity model (figure 3) has been derived from the 2.5D inversion of the CSEM data using the MARE2DEM code (Key, 2016). As observed on resistivity logs in the area, the resistivity model consists in shallow conductive sediments overlying deep resistive granitic basement.

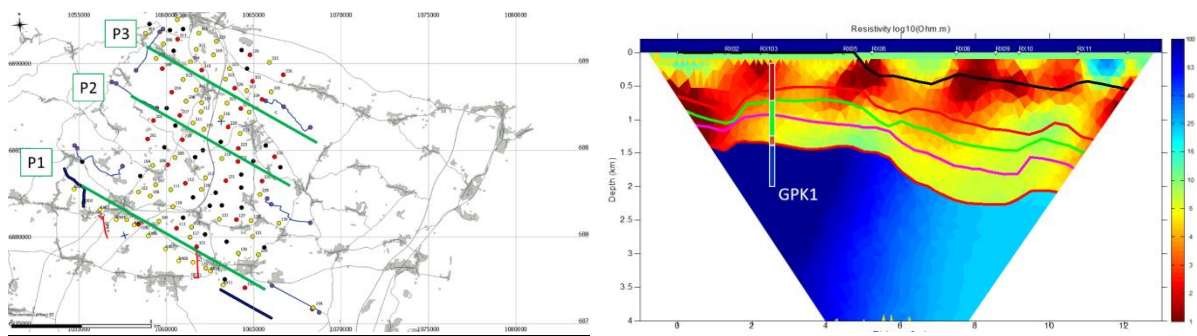

**Figure 3** *Left: 3D CSEM survey area showing the locations of CSEM receivers (circles) and transmitters (blue lines). Right: resistivity model obtained from the 2.5D inversion of the CSEM along profile P1.*

To extrapolate these results to the well and seismic line locations, a 3D CSEM inversion is ongoing. The initial step consists in running a 3D synthetic inversion study based on a simple resistivity model, using the exact acquisition geometry deployed in the field with both CGG RLM3D code and BRGM POLYEM3D code. The purpose of this synthetic inversion is to assess the resolution and sensitivity of our acquisition geometry to the key geothermal features (deep fractured networks, clays, fluids' salinity…) and set up the adequate inversion parameters. Actual data inversion will be subsequently performed to obtain a reliable 3D resistivity model of the area. This model will be finally coupled with the results of the seismic inversion based on the relationships established with the well/core analysis and synthetic resolution study.

**Conclusions**

Beyond this geothermal study, this research project is focusing on the challenges of integrating multi-scale and multi-physics data in a quantitative way. In particular, the non-linearity of the inverse problem is a key challenge. The use of Deep Neural Network approach may provide the ability of efficiently handling non-linear relationships between all input data to derive reservoir attributes. Application of such ML still requires careful quality control of the seismic or multi-physics data and the need to add more geological constrains.

**Acknowledgements**

The authors wish to thank the ADEME (French Agency of Energy and Environment) in the framework of the SIMGEO project. The authors acknowledge the GEIE EMC and ECOGI for their grateful help in this project.